\documentclass{aa}

\usepackage{multirow}
\usepackage{color}
\usepackage{xcolor}
\usepackage{graphicx}
\usepackage{hyperref}
\usepackage{pdflscape}
\usepackage{geometry}
\usepackage{tabularx} 
\usepackage{booktabs}
\usepackage{array} 
\usepackage{wasysym}
\usepackage[normalem]{ulem}

\graphicspath{{figures/}}

\hypersetup{
    pdftitle = {True obliquities distribution},
	pdfauthor = {A. M. Rossi et al.},
	colorlinks ={true},
	plainpages = {false},
	linkcolor={blue},
	citecolor={blue},
	urlcolor={blue}
}

\usepackage{cleveref}
\crefname{equation}{equation}{equations}
\crefname{chapter}{Appendix}{Appendices}
\Crefname{chapter}{Appendix}{Appendices}

\usepackage{physics}
\usepackage{diagbox}
\usepackage{booktabs}
\usepackage{xfrac}
\usepackage{units}

\usepackage{txfonts}

\def\Teff{T_{\mathrm{eff}}}

\def\vsini{\ensuremath{v\sin i_{\star}}}
\def\istar{\ensuremath{i_{\star}}}
\def\iorb{\ensuremath{i_\mathrm{orb}}}

\def\m2s2{\hbox{\,m$^{2}$\,s$^{-2}$}} 
\def\sini{\hbox{$\sin\istar$}}      
\def\cosi{\hbox{$\cos\istar$}}      
\def\Rsun{\hbox{$\mathrm{R}_{\odot}$}}

\def\Mearth{\hbox{$\mathrm{M}_\oplus$}}

\def\rstar{R_{\star}}     

\def\prot{P_{\mathrm{rot}}}

\begin{document} 

   \title{True spin-orbit obliquities distribution: data-driven confirmation of no clustering of misaligned planets}
   \titlerunning{True spin-orbit obliquities distribution}
   \authorrunning{A.~M.~Rossi et al.}
   \author{A.~M.~Rossi\inst{\ref{milanouniv}}, 
           M.~Rainer\inst{\ref{brera}},
           F.~Borsa\inst{\ref{brera}},
           S.~Facchini\inst{\ref{milanouniv}}
          }

   \institute{
Dipartimento di Fisica, Universit\`{a} degli Studi di Milano, Via Celoria
16, Milano, Italy \label{milanouniv}
 \and
INAF -- Osservatorio Astronomico di Brera, Via E. Bianchi 46, 23807 Merate (LC), Italy \label{brera}
             }
             \offprints{A.~M.~Rossi\\ \email{alessandromatteo.rossi@studenti.unimi.it}}

   \date{Received ; accepted }

 
  \abstract
   {True spin-orbit obliquities $\Psi$ offer valuable insights into the evolutionary history of exoplanetary systems. Previous studies have suggested that exoplanets tend to occupy either aligned or perpendicular orbits. However, recent research has indicated potential biases caused by the low sample, questioning whether this dichotomy would persist with a larger dataset. Simultaneously, a similar dichotomous behavior has been suggested for Neptune-sized planets.}
   {We aim to investigate the distribution of true spin-orbit obliquities $\Psi$ with an enlarged sample, looking for confirmation of the disputed dichotomy previously found, with a focus also on the obliquities of Neptunes.}
   {Starting from a sample of 264 projected obliquities $\lambda$, we homogeneously compute true obliquities $\Psi$ for 116 planets using the rotation period method. We combine them with 4 further values gathered from literature and we then study their distribution, also as a function of various star-planet system parameters.}
   {Our data-driven work based on 120 true obliquities $\Psi$ --- the largest sample to date --- strongly confirms the presence of a single cluster of aligned planets, followed by an isotropic distribution of misaligned planets with no preferred misalignment. This result is based on a uniform distribution of stellar inclinations $\istar$, for which non-uniformity could have biased previous interpretations of the arrangement of true obliquities. We confirm that Neptunians show a tentative dichotomous distribution with data available today, but its veracity needs confirmation with an enlarged sample, also because an anisotropic distribution of stellar inclination may be one of the factors hindering the real distribution.}
   {The future increase of the measured $\Psi$ sample over different planet types will allow better investigation of the relation between misalignment and system properties and help to depict a more comprehensive picture of the planetary evolution processes. 
   }


   \maketitle
%

\section{Introduction\label{sec:intro}}

The formation and evolution of planetary systems is one of the fundamental questions in modern astrophysics, as it provides key insights into not only the origin of our Solar System but also the diversity of thousands of exoplanets discovered in recent decades \citep{Raymond_2020,Kane_2021}.
It is still an open topic of debate, with many aspects yet to be fully understood \citep[e.g.][]{Mamajek_2009,Mordasini_2012,Turner_2014,Manara_2023}.
One of the key parameters that can provide valuable insights into planetary formation and evolution is the spin-orbit obliquity, the angle between the stellar rotation axis and the planet orbital plane. It is possible to derive the projected spin-orbit obliquity $\lambda$ --- as projected on the line of sight --- via the Rossiter-McLaughlin (RM) effect \citep{Rossiter_1924,McLaughlin_1924}, which is an apparent change in the radial velocity curve caused by a planet transiting in front of its host star's disk \citep[e.g.,][]{Triaud_2018}. 
Observational studies of projected spin-orbit obliquities in exoplanetary systems have revealed a diverse range of alignments, suggesting complex dynamical histories. While many hot Jupiters (HJs) exhibit low obliquities, indicating alignment with their host stars’ equatorial planes, a significant fraction display misalignments, including polar and even retrograde orbits (see \cite{Albrecht_2022} for a review). Trends have emerged linking obliquity distributions to stellar properties: planets around cool stars ($\Teff \leq 6250 \,\mathrm{K}$) tend to be well-aligned, whereas those orbiting hotter stars often show a broad range of misalignments, likely due to weaker tidal interactions \citep{Winn_2010,Albrecht_2012,Dawson_2014}. Additionally, multi-planet systems and smaller exoplanets generally exhibit lower obliquities \citep{Albrecht_2013,Winn_2015}, suggesting a different dynamical evolution compared to isolated hot Jupiters. Phenomena such as migration \citep{Kley_2012} and planet-planet scattering \citep{Rasio_1996, Weidenschilling_1996} are some of the possible explanations for exoplanets' tilted --- sometimes dramatically --- orbits.

Recent literature has focused on the distribution of true spin-orbit obliquities $\Psi$, which are obtained by detrending the values of $\lambda$ from the projection on the line of sight for those systems where it is possible to do this. 
Some authors have noted a dichotomy in the $\Psi$ distribution, with most planets clustering in aligned orbits ($0^\circ-41^\circ$), and a non-negligible fraction residing in nearly perpendicular orbits ($85^\circ-125^\circ$), leaving the intermediate spin-orbit inclinations empty \citep{Albrecht_2021}. Subsequent works have applied advanced statistical techniques and simulations to investigate how this dichotomy may arise from a biased sample. This bias, possibly caused by the stellar projected rotation velocities $\vsini$ or the stellar inclination $\istar$ selection \citep{Siegel_2023}, may obscure the true $\Psi$ distribution. Simulations indicate that the distribution of the projected spin-orbit angle $\lambda$ serves as a proxy for $\Psi$, suggesting a true spin-orbit distribution primarily composed of aligned planets, with a sparse population of misaligned ones and no significant clustering at obliquities near perpendicularity \citep{Dong_2023}. However, this has never been verified with actual data, as the sample of planets with measured $\Psi$ was too small. Other authors have also suggested a similar dichotomous trend in the projected obliquity distribution $\lambda$ of Neptune-sized planets \citep[e.g.,][]{Espinoza-Retamal_2024}, highlighting how compact systems tend to be aligned \citep{Razdom_2024}.

Among the $\sim6000$ confirmed exoplanets as of December 2024, only a few hundreds have a $\lambda$ value, and even fewer have a measure for the true obliquity $\Psi$ \citep{NEA,NEA_composite}. In this work, we compile the most up-to-date catalog of true obliquities by computing them using the \textit{rotation period method} or by collecting from literature. 
This paper is structured as follows. In \Cref{sec:bg} we detail the rotation period method. In \Cref{sec:catalog} we present our comprehensive catalog of exoplanets with $\lambda$ measures available in the literature. In \Cref{sec:simulation} we describe our Markov Chain Monte Carlo (MCMC) algorithm, created adopting widely accepted techniques, to compute true obliquities values $\Psi$ via the rotation period method. In \Cref{sec:results} we analyze the $\Psi$ distribution, and speculate about correlation with different parameters, such as stellar inclination and stellar age, focusing in particular on Neptune-sized planets. Finally, in \Cref{sec:discussion} we discuss possible mechanisms that could explain the distributions we infer. 

\section{From projected to true obliquities: the rotation period method\label{sec:bg}}

The RM effect provides information about the spin-orbit obliquity angle projected onto the plane of the sky $\lambda$, which corresponds to the line-of-sight difference between the host star spin axis and the planet orbital axis ($\istar$ and $\iorb$, respectively). The primary challenge to understand the geometry of planetary systems is to de-project this obliquity from the line of sight, to obtain the true obliquity $\Psi$. This can be achieved using the \textit{rotation period method}, a well established technique that constrains both the stellar inclination $\istar$ and $\Psi$ \citep[e.g.,][]{Brown_2012,Masuda_2020,Albrecht_2021,Bourrier_2023,Bowler_2023,Morgan_2024}. 
We briefly recall here the mathematical framework used to derive $\Psi$.

We can follow the approach of \citet{Fabrycky_2009} and derive an expression for the cosine of the true obliquity
\begin{equation}
    \cos\Psi = \sini \cos\lambda\sin \iorb + \cosi\cos \iorb
    \label{eq:cospsi}
\end{equation} 
which depends on the stellar inclination $\istar$, the sky-projected spin-orbit angle $\lambda$, and the orbital inclination $\iorb$. 
Since the RM effect occurs during a planetary transit, one could be tempted to wrongly assume an orbital inclination $\iorb\approx 90^\circ$, thus simplifying the previous equation by dropping the last term. In reality several HJs show inclinations lower than $85^\circ$ (e.g. TOI-1431b/MASCARA-5b shows an orbital inclination $\iorb = 80.3^{+0.18}_{-0.17}{}^\circ $\citep{Stangret_2021} whereas HAT-P-56b has $\iorb=82.6^{+0.7}_{-0.6}{}^\circ$\citep{Zhou_2016}). Considering that a drop of $\sim10^\circ$ around $90^\circ$ produces a variation of $17\%$ in the value of the cosine, properly accounting for them is mandatory for a correct inference. 

Neglecting the star's differential rotation, we can define the equatorial velocity of the host star
\begin{equation}
    v_{\mathrm{eq}} = \frac{2\pi\rstar}{\prot}
    \label{eq:veq}
\end{equation}
and with this we can express $\sini$ as
\begin{equation}
    \sini = \frac{v_{\mathrm{eq}}\,\sini}{v_{\mathrm{eq}}} = \frac{\prot\,\vsini}{2\pi \rstar}
    \label{eq:sini}
\end{equation}
where $\prot$ is the stellar rotation period, $\vsini$ is the stellar rotational velocity projected along the line-of-sight, and $\rstar$ is the stellar radius. A nice visualization of this geometry is given in Fig. 1 of \citet{Albrecht_2022}.

Theoretically, the stellar inclination $\istar$ could be constrained also using different approaches. Late-type stars can have their $\istar$ determined through asteroseismology \citep{Gizon_2003,Ballot_2006} in a complementary fashion with respect to the RM, also providing higher precision. However, this technique is possible only for bright stars, with its reliability strongly suffering biases when the signal-to-noise ratio of the power spectra is low \citep{kamiaka18}.
Another method of determining $\lambda$ and $\istar$ is to analyze the possible asymmetry of transit light curves, caused by gravity darkening \citep{barnes2009}.

For the sake of homogeneity, we opted to use \cref{eq:cospsi} for the creation of our sample where possible. When the stellar rotation period --- or other quantities of \cref{eq:cospsi} --- is not available, we used literature values of $\Psi$ estimated through other methods.


\section{Creating a catalog: sample selection\label{sec:catalog}}

We started the creation of our obliquities sample with the 235 planets with available $\lambda$ values in the TEPCat catalog\footnote{https://www.astro.keele.ac.uk/jkt/tepcat/obliquity.html} \citep{Southworth_2011} as of December 2024.
We then extended the sample by adding 21 additional planets from the NASA Exoplanet Archive\footnote{https://exoplanetarchive.ipac.caltech.edu/} \citep{NEA,NEA_composite,NEA2025}, hereafter referred to as NEA. 
When multiple $\lambda$ values were available for the same planet, we considered the most recent, precise, and comprehensive sources as more eligible. 
 
A bibliographic search allowed us to retrieve 18 more $\lambda$ values that were still not present in the catalogs, probably because they were recently published. These values were obtained for the planets TOI-2145b \citep{Dong_2024}, TOI-2119b \citep{Doyle_2024}, TOI-1694b \citep{Handley_2024}, TOIs 558b, 2179b, 4515b, and 5027b \citep{Espinoza-Retamal_2024}, TOI-1259Ab \citep{Veldhuis_2025}, TOI-1759Ab \citep{Polanski_2025} TOI-2364b \citep{Tamburo_2025}, TOI-3714b and TOI-5293Ab \citep{Weisserman_2025}, K2-237b \citep{Zak_2025b}, as well as HAT-P-50b, Qatar-4b, TOI-2046b, WASP-48b and WASP-140b \citep{Zak_2025}. 
This resulted in a total of 274 individual planets (orbiting 265 stars) with known projected obliquity $\lambda$. For systems hosting multiple planets, we restrict our analysis to a single planet, as mutual inclinations are beyond the scope of this study. So we ended up with 264 planets with known projected obliquity $\lambda$, 71 of which also have a published true obliquity $\Psi$. The only missing $\lambda$ is Kepler-408b as we were only able to retrieve its $\Psi$ value from \citet{Kamiaka_2019}.

We aimed at uniformly computing $\Psi$ from the literature values of $\lambda$ for all the planets of our sample. To this end, we needed reliable --- and possibly homogeneous --- estimates of stellar radii ($\rstar$), stellar rotational periods ($\prot$), and stellar projected rotational velocities (\vsini) to derive stellar inclinations $\istar$ (see \cref{eq:sini}), and subsequently $\Psi$, as shown in \cref{eq:cospsi}.

\subsection{Stellar rotation periods}
The stellar rotation period was the most limiting parameter in our work: after an extensive bibliographic search, we were able to recover the rotation periods $\prot$ only for 120 host stars. This is due to the need to observe long photometric time series (up to several months of observations) to obtain $\prot$. The stellar rotation period is derived from the modulation of the stellar brightness due to the presence of magnetically active regions on the stellar rotating surface: non-active stars (e.g., hot A-F type stars) may lack any feature that allows us to derive the stellar rotation, while very active stars (e.g. young stars, or M-type stars) may present a large number of spots and/or faculae that may drive to misidentify $\prot$. Additionally, a partial or inhomogeneous observing window may cause an alias of the $\prot$ being identified as the true rotational period \citep{Basri_2020}. Also the presence of differential rotation may lead to incorrect $\prot$ values \citep{Aigrain_2015}.

Regarding the literature values of our sample, one host star exhibited multiple $\prot$ measurements. Three more had the rotation period either reported without uncertainties or only as an upper limit. In detail:

\begin{itemize}
    \item WASP-60 \citep{Mancini_2018}: two compatible $\prot$ values were provided ($P_\mathrm{rot,1} = 31.8\pm1.9\,\mathrm{days}$ and $P_\mathrm{rot,2} = 34.8\pm 2.7\,\mathrm{days}$); we conservatively assigned the semi-difference of the maximum range as the uncertainty for the central value ($\prot = 33.3 \pm 3.8\,\mathrm{days}$);
    \item HD 17156 \citep{Fischer_2007}: the rotation period was provided without uncertainties ($\prot=12.8\,\mathrm{days}$), so we supplemented it with a conservative $10\%$ uncertainty, a typical uncertainty for F- and G-type stars \citep{Santos_2021};
    \item WASP-180A \citep{Temple_2019}: literature provided only an upper limit on the rotation period ($\prot <3.3\,\mathrm{days}$), due to the presence of its co-moving companion, WASP-180B, which complicates measurements. We allowed the value to span values from the upper limit down to the break-up period of $0.1\,\mathrm{days}$.
    \item KELT-18 \citep{Rubenzahl_2024}: similarly to the previous case, literature provided only an upper limit. We allowed the value to span from this upper limit to the break-up period of $0.4\,\mathrm{days}$.
\end{itemize}

\subsection{Stellar radii}

Stellar radii may be computed with very different methods: the most precise measurements are those obtained using interferometry and asteroseismology, or for eclipsing binary stars \citep{2018Moya}. The latter are absent from our sample, while interferometry and asteroseismology may be used only on a limited sample of bright targets. Stellar radii may be derived also from photometric data: in this case, a good estimate of the distance and the interstellar extinction is crucial. Photometric radii are the most commonly found in literature, but the data may be taken with many different instruments.

In this work, we decided to use the radii given in the Gaia DR3 Astrophysical Parameters Archive\footnote{\url{https://doi.org/10.17876/gaia/dr.3/43}}, to ensure the most homogeneous and precise set possible. Two values of $\rstar$ are present in the archive, the GSP-Phot and the FLAME $\rstar$. The first are computed using the state-of-the-art Gaia parallaxes, and the photometric and spectroscopic Gaia data (the low-resolution BP/RP spectra and the apparent G magnitude), while the latter are a refinement built on both the photometric and spectroscopic Gaia data \citep{GaiaDR3}. We used the GSP-Phot values for our sample. We however note that the two sets of values are in very good agreement.

Our sample consists of 265 planets: in 22 cases, we were unable to retrieve Gaia information, and as such we had to use literature values. However, the final sample will have only 12 true obliquities computed without Gaia radii ($10\%$), because of the shortage of parameters needed to compute \cref{eq:cospsi}, hence granting a large homogeneity.

In the end, between Gaia DR3 and literature search, we were able to estimate the stellar radii for all the 265 stars in our sample.

\subsection{Projected rotational velocities and orbital inclinations}

Projected rotational velocities ($v\sin i_\star$) were taken from the same analyses providing the sky-projected obliquity $\lambda$, whenever available. These values often combine spectroscopic $v\sin i_\star$ measurements with the additional constraints from the Rossiter–McLaughlin effect, which can lead to improved precision, especially at low $v\sin i_\star$, compared to estimates derived from CCFs alone. We were able to retrieve 264 projected rotational velocities \vsini.

Similarly to $\vsini$, we preferably gathered orbital inclinations from the same paper originating information about $\lambda$, to ensure homogeneity, for a total of 263 \iorb.

In the end, our sample of 265 planets --- 71 of which had a $\Psi$ measure available --- presented 264 projected obliquities $\lambda$, 265 stellar radii $\rstar$ (243 of which from GAIA DR3 Astrophysical Parameters), 264 \vsini, and 120 rotation periods $\prot$. We were able to obtain the complete set of all the four values for 119 planets. Kepler-408 b is missing both $\lambda$ and \vsini, but we have information about its star's rotation period.

\subsection{True obliquities $\Psi$ from literature}

Our literature search turned out 8 more $\Psi$ for planets where we were not able to recover all the information needed for the rotation period method, bringing our total sample of planets with information about $\Psi$ --- or about the parameters to infer it --- up to 127 $\Psi$. 

\citet{Siegel_2023} proved that gravity darkening (GD) is biased towards perpendicular orbits, so we decided to exclude 7 measurements from our sample, because their obliquity, either projected or true, was extracted via GD. For four out of this seven we had only information about the true obliquity (we were missing all the four parameters needed to apply the rotation period method): HATS-70 b \citep{Zhou_2019}, KELT-17 b \citep{Zhou_2016}, KELT-19 b \citep{Kawai_2024}, Kepler-1115 b \citep{Barnes_2015}.
For the remaining three the literature provides all the parameters required for our analysis: KOI-13 b \citep{Howarth_2017}, Kepler-462 b \citep{Ahlers_2015}, and MASCARA-1 b \citep{Hooton_2022}.

It is worth noting that, within the limited sample of planetary obliquities measured via GD, some systems also have obliquity determinations from the RM effect, which yield results in close agreement (for example KELT-9 b \citep{Borsa_2019,Ahlers_2020,Stephan_2022} or MASCARA-4 b \citep{Ahlers_2020,Zhang_2022}).

With this final modifications we have no GD measurements in our sample, and a total of 116 planets for which the rotation period method is applicable, supplemented by 4 more true obliquities $\Psi$ found in literature: HD 89345 b \citep{Bourrier_2023}, KELT-9 b \citep{Stephan_2022}, Kepler-408 b \citep{Kamiaka_2019}, and WASP-131 b \citep{Doyle_2023}. The first true obliquity was detected via RM combined with a PDF for $\istar$ from \citet{VanEylen_2018}, the second via Doppler tomography (DT), the third had the stellar inclination determined via asteroseismology \citep[e.g.,][]{Ulrich_1986, Aerts_2010}, and the fourth via the Reloaded Rossiter-McLaughlin \citep[RRM][]{Cegla_2016} technique.

An extract of our dataset is available in \cref{tab:data} and in machine-readable form at CDS in its complete form. It also includes the results of our analysis, detailed in the following section.

\section{Inferring true obliquities\label{sec:simulation}} 

\begin{figure}[h]
    \centering
    \includegraphics[width=\linewidth]{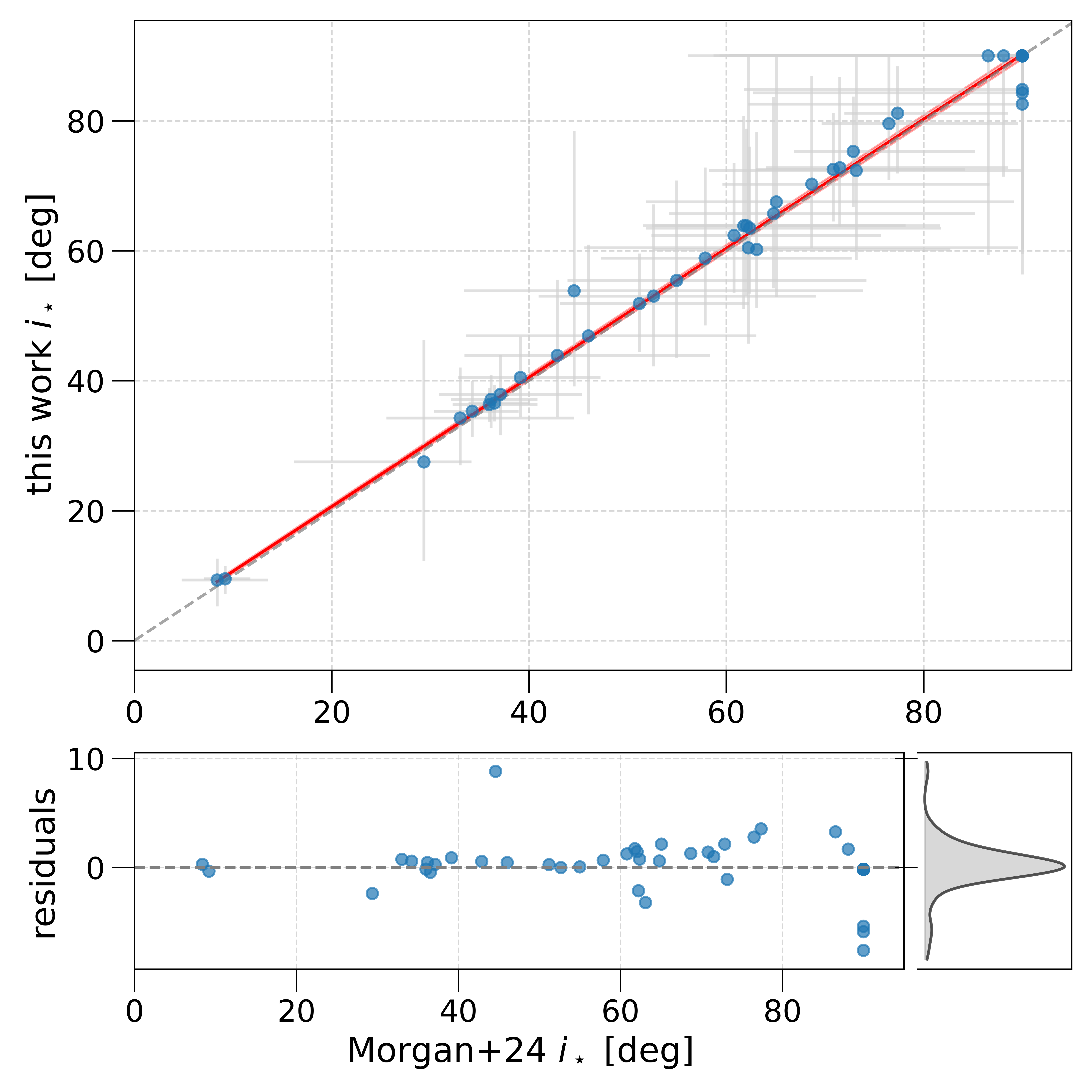}
    \caption{Comparison between our stellar inclinations $\istar$ and the ones computed by \citet{Morgan_2024}, starting from the same dataset. The gray dashed lines represent the $y=x$ relation, whereas the solid red line represents the linear fit, and the shaded red region represents the $68\%$ confidence interval. Residuals are computed with respect to the fit relation. Uncertainties in the residuals panel are omitted to ensure a better readability.}
    \label{fig:simVSmorgan}
\end{figure}

We computed the true obliquities using a robust MCMC algorithm. Firstly, we computed the stellar inclinations exploiting \cref{eq:sini}, accounting for the inherent correlation between $v_{\mathrm{eq}}$ and \vsini  highlighted by \citet{Masuda_2020}, mandatory for every inference involving these two values (e.g. \istar and $\Psi$ as in this work). The approach we adopted also follows the methodology detailed in Appendix A of \citet{Bowler_2023}. It is performed assuming uniform priors on $\vsini$, $\rstar$, $\prot$, along with an isotropic ($\sini$) prior on the stellar inclination. We tested our MCMC algorithm to extract stellar inclinations \istar on \citet{Morgan_2024} dataset, yielding great compatibility among $\istar$ results (p-value = $1.0$), as shown in \cref{fig:simVSmorgan}. 
Lastly, by incorporating a uniform prior on the projected obliquity $\lambda$, we extracted true obliquities $\Psi$\footnote{The $\Psi$ values in \cref{tab:data} assume a conventional orientation for the stellar spin axis ($\istar \leq 90^\circ$). This is a reasonable approximation considering that the median spacing between two degenerate solutions of \cref{eq:cospsi} is $0.26\sigma$.} using \cref{eq:cospsi}.
\begin{figure}[h]
    \centering
    \includegraphics[width=\linewidth]{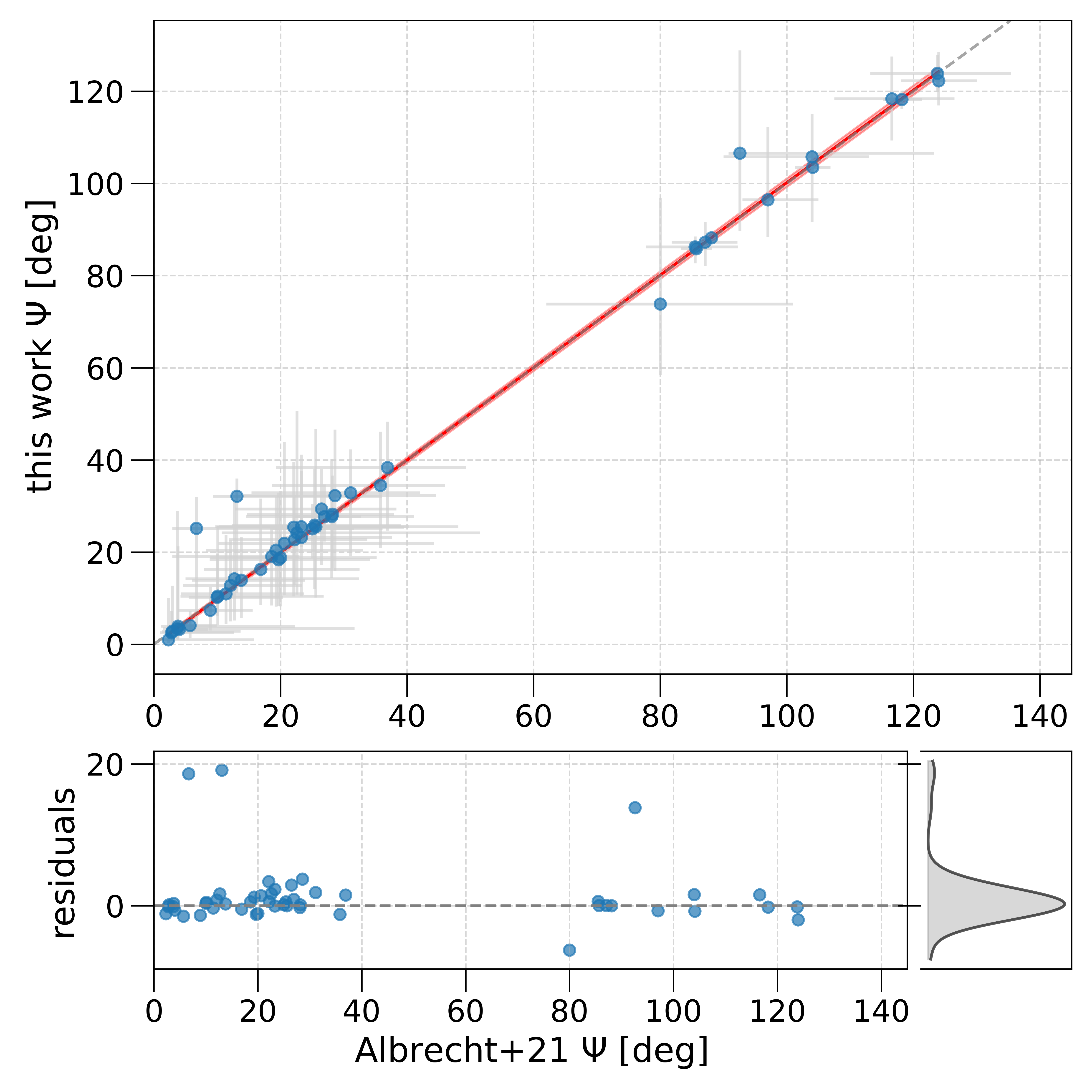}
    \caption{Comparison between our true obliquities and the ones computed by \citet{Albrecht_2021}. Starting from the same dataset, we achieve an excellent agreement within the $68\%$ confidence interval of the linear fit. The light gray dashed line represents the $y=x$ relation. The linear correlation is guaranteed by a $\chi^2$ test yielding $p=1.0$, allowing the rejection of the null hypothesis that the data are not linearly correlated. Residuals are computed with respect to the fit line. Uncertainties in the residuals panel are omitted to ensure a better readability. The presence of outliers is due to our inability to recover the $\iorb$ values used by \citep{Albrecht_2021}, but also highlights their importance for a correct inference.}
    \label{fig:simVSalbrecht+21}
\end{figure}

\begin{figure}[h]
    \centering
    \includegraphics[width=\linewidth]{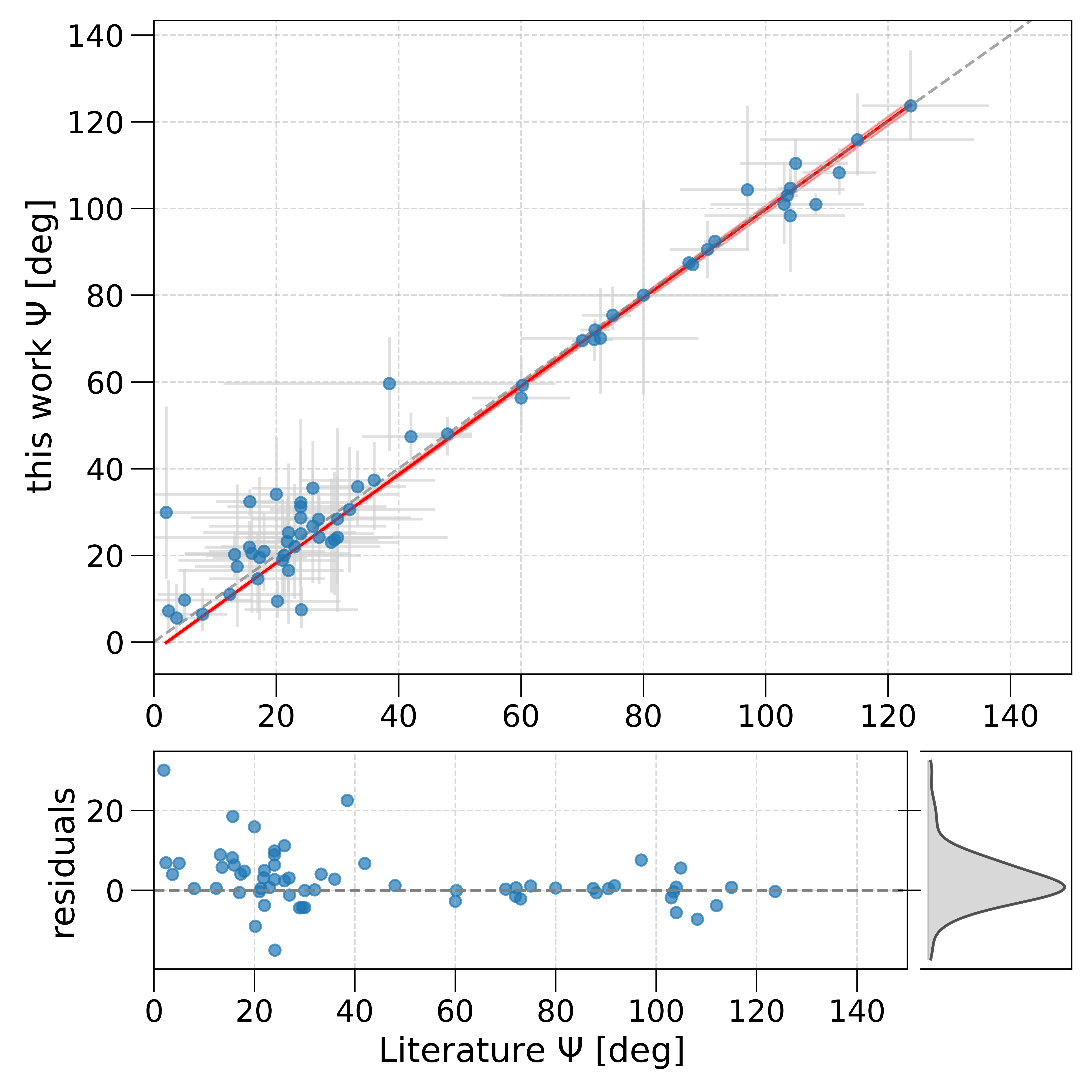}
    \caption{Comparison between the true obliquities computed in this work and correspondent literature values, aside from \citet{Albrecht_2021}, where available. Aesthetics follow the style of \cref{fig:simVSalbrecht+21}. The plot is made assuming the conventional stellar inclination ($\istar \leq 90^\circ$).}
    \label{fig:simVSliterature}
\end{figure}

The task of extracting true obliquities from projected obliquities has been previously undertaken by several authors, including \citet{Albrecht_2021}. 
We tested our algorithm on their dataset of planets with computable ($\rstar$, $\prot$, $\vsini$ and $\lambda$) true obliquities $\Psi$, obtaining excellent agreement between the two results, as illustrated in \cref{fig:simVSalbrecht+21}. 
In 63 cases our computed true obliquities had a counterpart in studies other than \citet{Albrecht_2021}, 4 of which were only upper limits, namely HAT-P-36 b \citep{Mancini_2015}, Kepler-17 b \citep{Desert_2011}, WASP-43 b \citep{Esposito_2017}, and TOI-1136 d \citep{Dai_2023}. We compared the results of the MCMC algorithm also with 59 of these values finding good agreement: a Pearson test yielded a 0.99 correlation rank with a p-value order of magnitudes lower than the threshold of 0.05 to reject the null hypothesis of data being uncorrelated. As radii values were mostly taken not from the work from which we gathered $\lambda$ and \vsini (\cref{fig:simVSliterature}), and since literature $\Psi$ values do not represent an homogeneously derived sample, we obtained p-value lower than the aforementioned threshold while performing an ordinary linear regression. Hence, we can conclude that our inference method is consistent with other works when we apply it on their dataset \citep{Albrecht_2021,Morgan_2024}, and it yields coherent results with literature values. The four excluded upper limits are compatible with our results.

Therefore, we decided to adopt literature values for the true obliquity only if one of the parameters mentioned in \Cref{sec:catalog} was missing, preventing us from calculating that obliquity.

\section{Results\label{sec:results}}

\begin{figure*}[h]
    \centering
    \includegraphics[width=.9\linewidth]{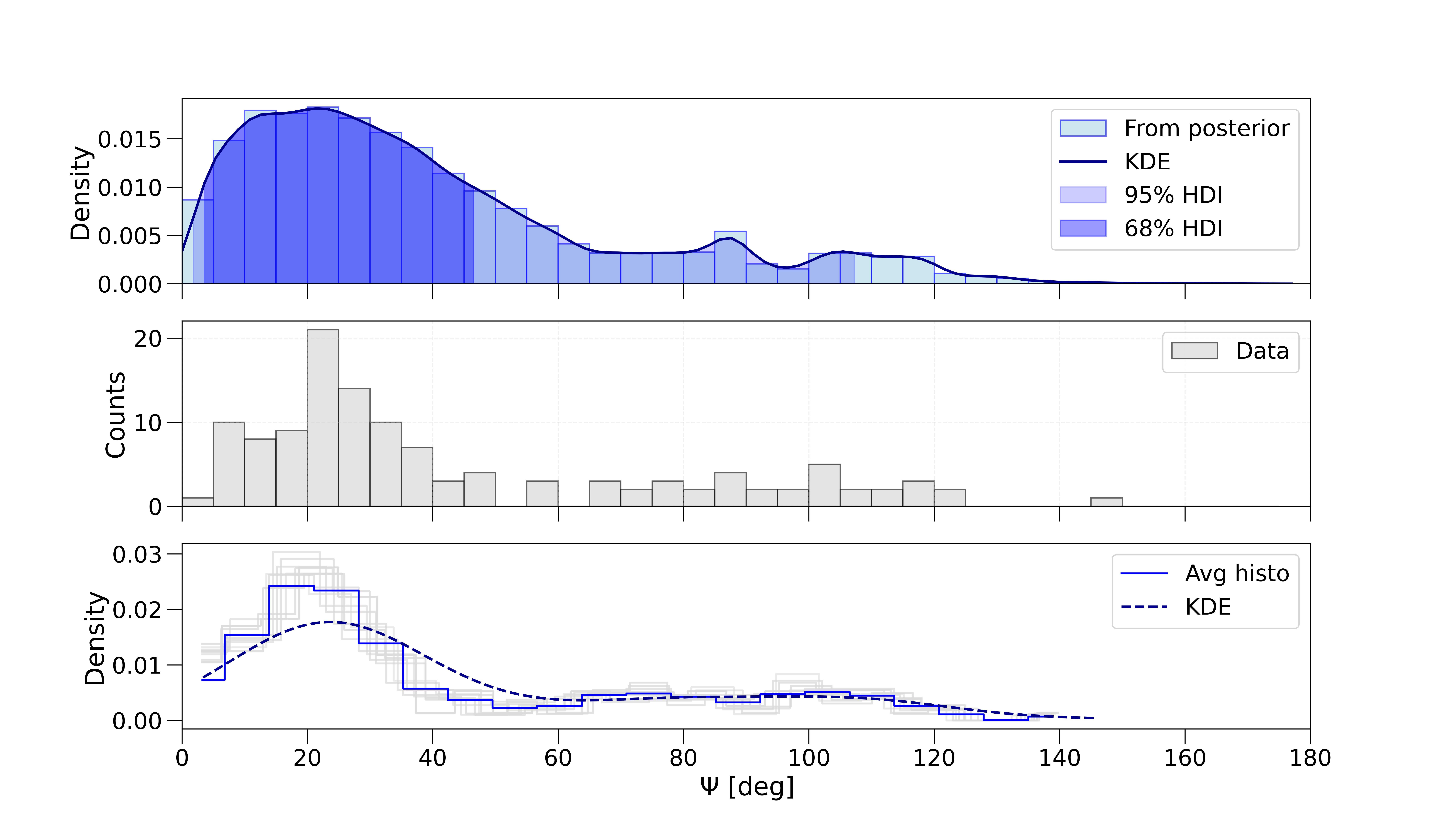}
    \caption{Distribution of the 120 true obliquities present in our sample. In the top panel, $\Psi$ values are shown as a superposition of their normalized posteriors, fitted with a Kernel Density Estimation (KDE) with Gaussian kernel and Silverman's bandwidth (other bandwidth selections yielded either similar results or overfitted graphs). The shaded blue and light blue regions represent $68\%$ and $95\%$ highest density intervals (HDI), respectively. The middle panel displays a simple histogram of true obliquities. The lower panel addresses two key issues: (i) the arbitrariness of binning; (ii) the apparent valley around $90^\circ - 100^\circ$. To account for these, 100 histograms with 20 bins were generated, with $20\%$ variability on their boundaries. The average histogram is represented by the solid blue line, while the dashed blue line represents the KDE.}
    \label{fig:psi_distribution}
\end{figure*}

In \cref{fig:psi_distribution} we show the distribution of our 120 $\Psi$ values, that doubles the observational sample of 57 planets presented in \citet{Albrecht_2021}.
Only a cluster of aligned planets is present, with no significant density around perpendicular orbits. Our results thus confirm with observational data what was previously supposed by \citet{Siegel_2023} and \citet{Dong_2023}.

We examined the robustness of our analysis, particularly concerning the disputed apparent lack of planets in the $90^\circ-100^\circ$ range, both in the posterior probability histogram and in the distribution of central values (\cref{fig:psi_distribution}, top and middle panels). After testing different kernels and bandwidth selections \citep{Scott_1992,Silverman_2018} --- which had no significant impact on the overall shape --- we adopted an averaged histogram with variable binning (bottom panel). Specifically, we computed 100 histograms with 20 bins, introducing $20\%$ bin variability to mitigate both overfitting and oversmoothing. This approach reduced the apparent valley in the perpendicularity range. Finally, we computed a Kernel Density Estimation (KDE) on the now smoother average histogram, which presented a single prominent peak for aligned orbits.

We performed a Kolmogorov-Smirnov (K-S) test with the null hypothesis that the $\Psi$ distribution from \citet{Albrecht_2021} is the same as the distribution of our dataset, using a significance level of $p = 0.05$. The resulting p-value did not allow us to reject the null hypothesis. This does not hinder our conclusions, because small sample sizes can lead the K-S test to not rejected the null hypothesis even though two distributions differ: an alternative can be an Anderson-Darling test \citep[A-D][]{ADtest1,ADtest2}, which places more weight on the tails of the distribution without relying only on the maximum spacing among them \citep{ADvsKS1, ADvsKS2, ADvsKS3}. The A-D test rejected the null hypothesis of the two sample being drawn from the same distribution with a p-value of $p=8\cdot 10^{-4}$. However, we note that this result could be biased by an overall small sample, of barely more than a hundred planets. Hence, considering that \citet{Albrecht_2021} had suggested that true obliquities cluster on aligned and perpendicular orbits, we performed a dip-test for uni-modality \citep{diptest} on their sample. As expected, we can reject the null hypothesis of their sample being drawn from a uni-modal distribution at the $1\%$ level, whereas our sample has a p-value of $0.91$. These facts highlight how our $\Psi$ sample differs from \citet{Albrecht_2021} in terms of modality of the distribution, being close to what more recent literature expects \citep{Dong_2023}. It should be noted --- and it is reasonable to expect --- that with a larger sample size also the K-S test would reject the null hypothesis.

\begin{figure}[h]
    \centering
    \includegraphics[width=\linewidth]{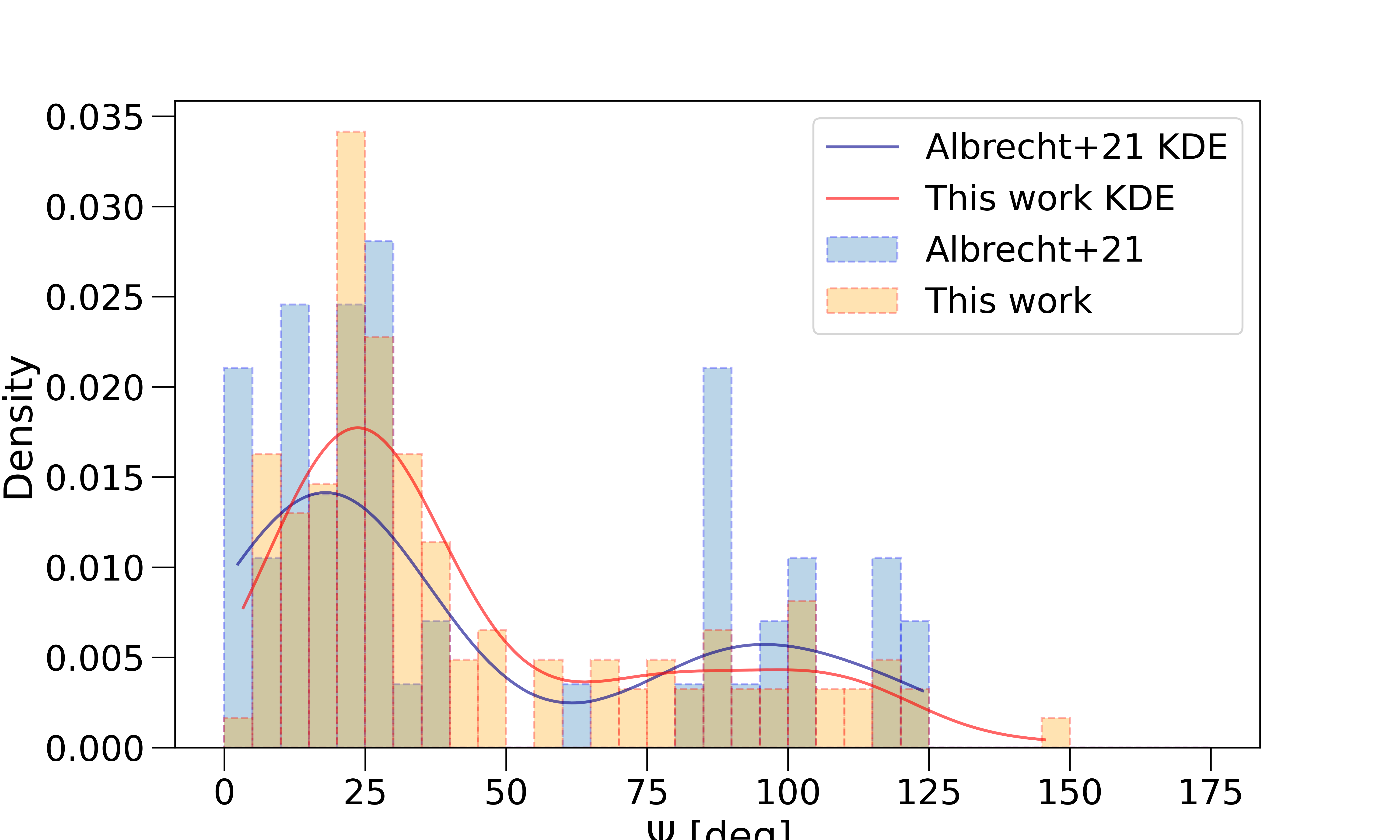}
    \caption{Comparison between the dataset of $\Psi$ from \citet{Albrecht_2021} and that of this work. An A-D test provided evidence that the two distributions significantly differ, along with a dip-test for uni-modality proving that \citep{Albrecht_2021} true obliquities distribution is consistent with a multi-modal distribution while our sample of true obliquities is consistent with a uni-modal distribution. KDEs are plotted for ease of visualization.}
    \label{fig:compareKDE}
\end{figure}

\subsection{Stellar inclination distribution}

\begin{figure}[h]
    \centering
    \includegraphics[width=\linewidth]{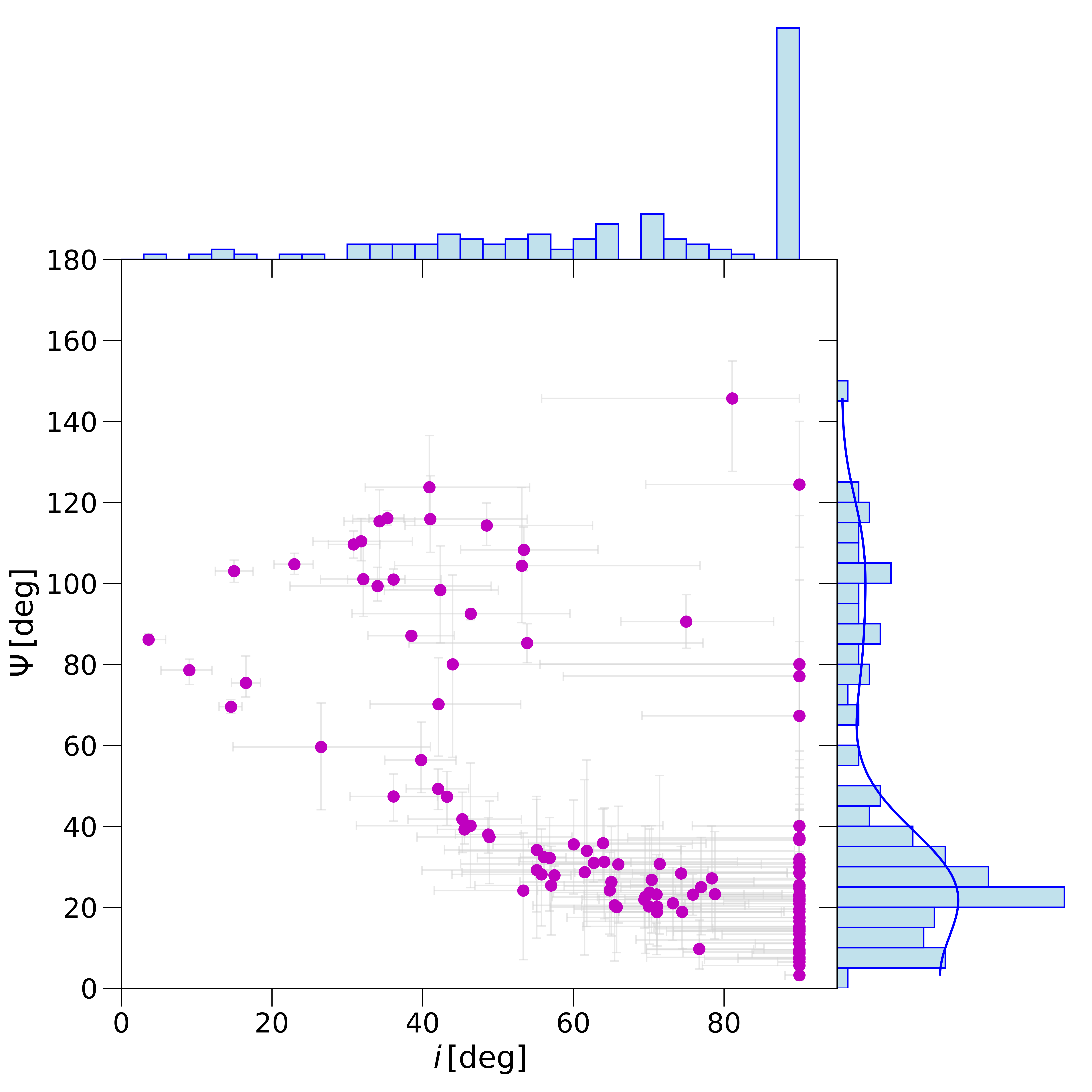}
    \caption{True spin-orbit obliquities of our sample of 118 planets in respect to the stellar inclination. Marginal histograms show a peak for perpendicular stellar inclinations $\istar$ (top) and a peak for aligned orbits followed by a near-uniform distribution for misaligned orbits (right). Marginal histogram for $\Psi$ is binned every $5^\circ$ and is fitted with a KDE with Gaussian kernel and Silverman's bandwidth.}
    \label{fig:psiVSistar_sim}
\end{figure}
\begin{figure}[h]
    \centering
    \includegraphics[width=\linewidth]{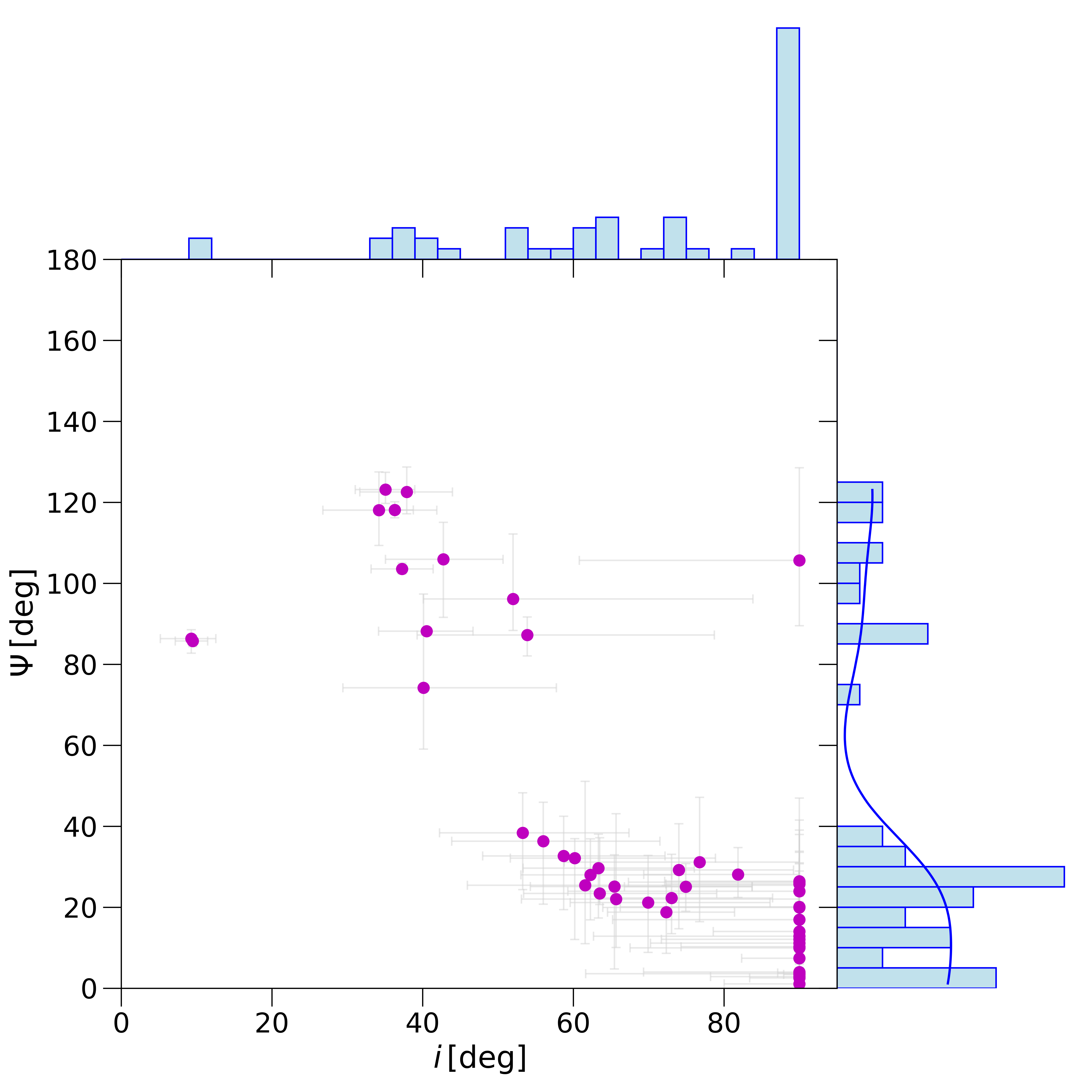}
    \caption{True spin-orbit obliquities from \citet{Albrecht_2021} with respect to stellar inclination. We were able to re-compute a total of 51 $\Psi$ values, thanks to the availability of the quantities in \cref{eq:cospsi}. The marginal histogram on the right (binned every $5^\circ$ and fitted with a KDE using a Gaussian kernel and Silverman’s bandwidth) highlights a dichotomy in true obliquities, while the top marginal histogram emphasizes the non-uniformity of stellar inclinations $\istar$.}
    \label{fig:psiVSistar_albrecht}
\end{figure}
\begin{figure}[h]
    \centering
    \includegraphics[width=\linewidth]{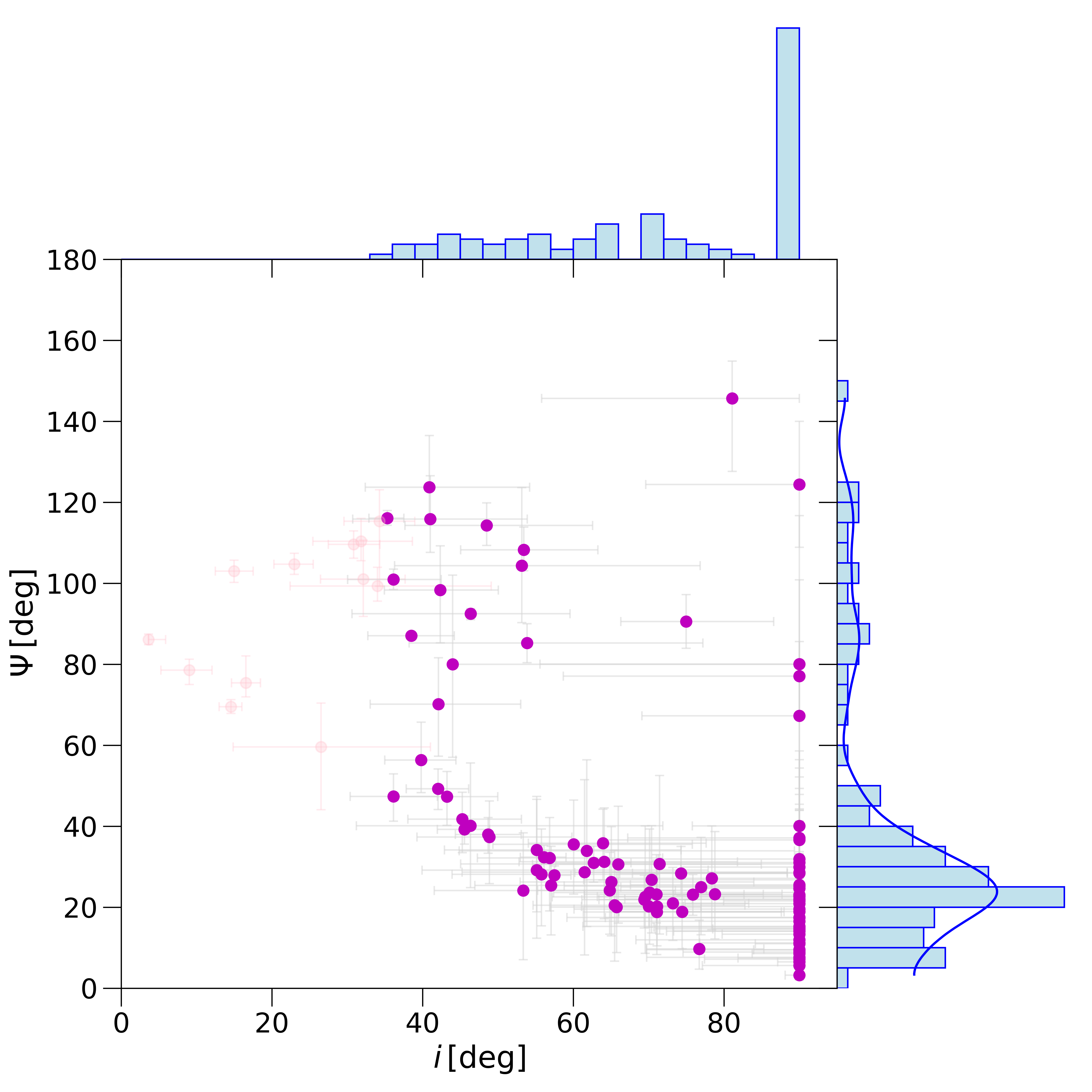}
    \caption{Re-creation of the apparent dichotomy found by \citet{Albrecht_2021} in respect to stellar inclination. In this figure, following the same fashion of the previous, we approximately recreated an apparent dichotomy for true obliquities $\Psi$ by removing only 12 planets from the 118 available. This result underscores the importance of uniformity on stellar inclinations $\istar$. Low stellar inclinations $\istar$ are key to obtain uniformity on misaligned values of $\Psi$.}
    \label{fig:psiVSistar_tagliato}
\end{figure}

Our algorithm also provided information about the stellar inclination distribution within our sample, shown in \cref{fig:psiVSistar_sim}. This was possible only for the 116 systems for which we computed \cref{eq:cospsi}, along with HD 89345 b and WASP-131 b for which we were able to retrieve information about \istar, for a total of 118 planets. We investigated $\istar$ as the possible bias that led \citet{Albrecht_2021} to identify the dichotomy shown in the right marginal histogram of \cref{fig:psiVSistar_albrecht}, as suggested by \citet{Dong_2023} and \citet{Siegel_2023}. By re-computing stellar inclinations and true spin-orbit obliquities for the 51 planets with $\rstar$, $\prot$, $\vsini$, $\lambda$ available in their dataset, we identified two key factors: (i) the sample size and (ii) the diversity of stellar inclinations. Our sample is significantly larger ($+127\%$) and our stellar obliquities are more evenly distributed across the admissible range $0^\circ-90^\circ$. We confirmed this trend in \cref{fig:psiVSistar_tagliato}. By removing stellar inclinations lower than $35^\circ$ --- a reasonable threshold when comparing the $\istar$ distributions between \cref{fig:psiVSistar_sim} and \cref{fig:psiVSistar_albrecht} --- we were able to reproduce an apparent dichotomy while minimizing data loss (only 12 planets were removed). 

Considering that leading theories of planetary formation predict that planets form in approximately aligned orbits within an approximately aligned protoplanetary disk \citep[e.g.,][]{Ida_2004,Armitage_2010}, it is reasonable to expect that most planets will remain on aligned orbits. This expectation holds particularly for perpendicular stellar inclinations. Conversely, the detection of an exoplanet --- and a subsequent RM measure --- can only occur if the planet transits in front of its host disk. Thus, stellar inclinations closer to a pole-on configuration ($\istar\sim 0^\circ$) imply the planet will be on a misaligned orbit \citep{Campante_2016}. This, combined with a large sample size, can manifest a noticeable peak on aligned orbits with a more uniform spread at other angles.

\subsection{True obliquities vs. stellar age}

Since true obliquity can serve as a proxy for the dynamical evolution of an exoplanetary system, we investigated whether it correlates with stellar age. Stellar age is notoriously a challenging parameter to measure with high precision \citep{Soderblom_2010}. After an extensive literature search, we identified the stellar age for all the 120 stars for which we had a true obliquity $\Psi$ available, except one --- K2-105 --- totaling 119. 

\begin{figure}[h]
    \centering
    \includegraphics[width=\linewidth]{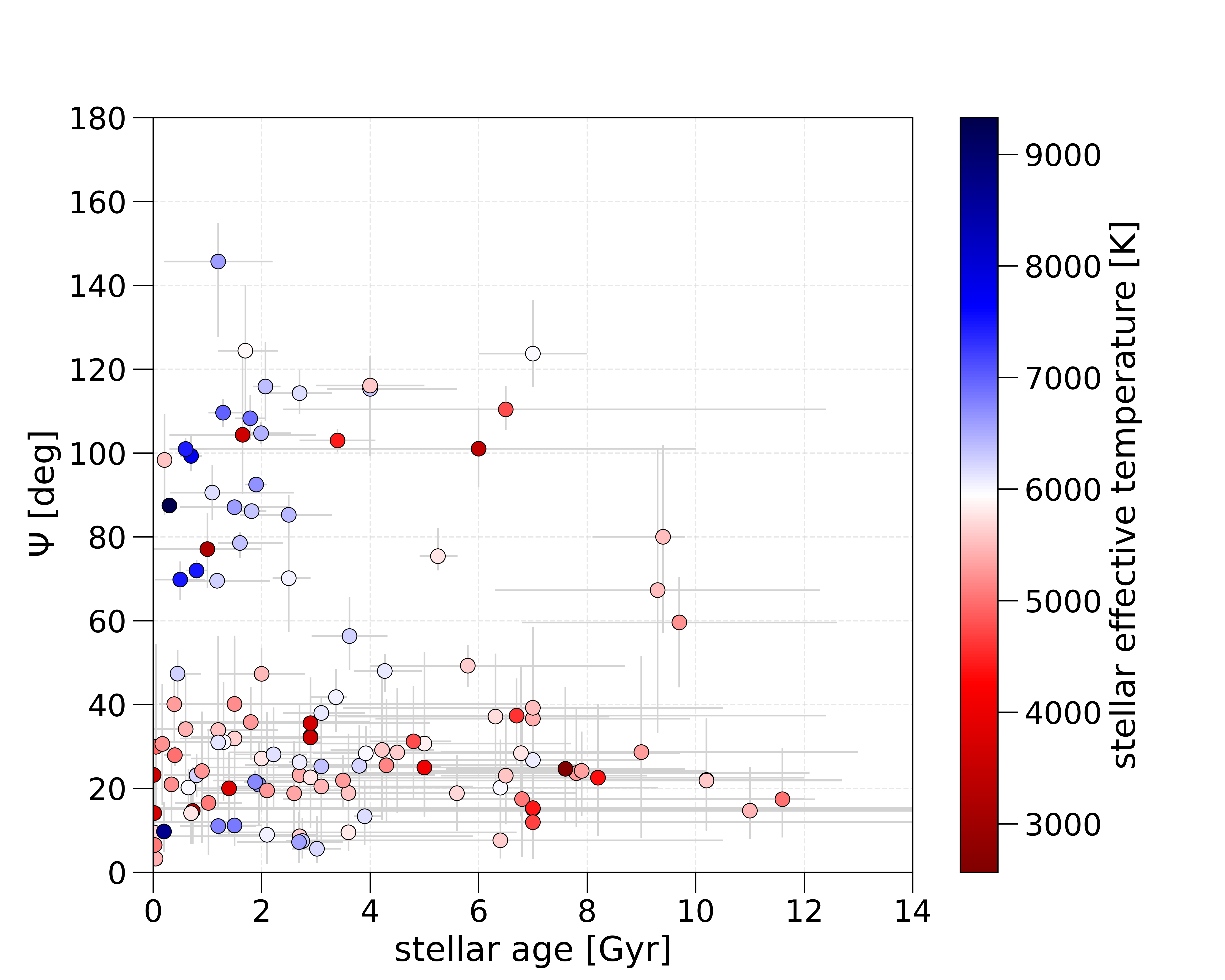}
    \caption{Distribution of obliquities $\Psi$ with respect to stellar age for our sample. 119 planets are plotted, as K2-105 is excluded because its age estimate is not available in the literature.}
    \label{fig:psiVSage}
\end{figure}

In \cref{fig:psiVSage} we analyze the relation between $\Psi$, the stellar age and the stellar effective temperature. Defining misaligned planets as those with an obliquity greater than $41^\circ$, angle beyond which there can be von Zeipel-Kozai-Lidov cycles \citep[ZKL][]{vonZeipel_1910,Kozai_1962,Lidov_1962}, we find a predominance of misaligned planets among stars younger than $5\,\mathrm{Gyr}$, with minor exceptions. The existence of two subpopulations in \cref{fig:psiVSage} is evident: one consisting of young and aligned planets, and the other of young and misaligned planets. Older misaligned planets do not contribute significantly, primarily due to the large uncertainties associated with their age estimates. Additionally, aligned planets systematically exhibit lower host star temperatures compared to misaligned ones. This latter interpretation, however, is likely biased by the stellar age itself, as higher temperatures are not expected for old stars. 

We are not able to extract any conclusion regarding the relation between obliquity and stellar age, except for the --- already known --- fact that planets tend to damp obliquity overtime via several mechanisms (or a combination of them), such as tidal dissipation \citep{Ogilvie_2014,Lin_2017}, von Zeipel-Kozai-Lidov cycles in presence of a distant perturber \citep{vonZeipel_1910,Kozai_1962,Lidov_1962,Katz_2011}, and secular interactions \citep{Saillenfest_2019}. However we noted a cluster of older and misaligned planets warranting investigation.

\subsection{A focus on Neptunian planets} \label{subsec:psiage}

\begin{figure*}[h]
    \centering
    \includegraphics[trim=10pt 0pt 10pt 0pt, clip, width=\linewidth]{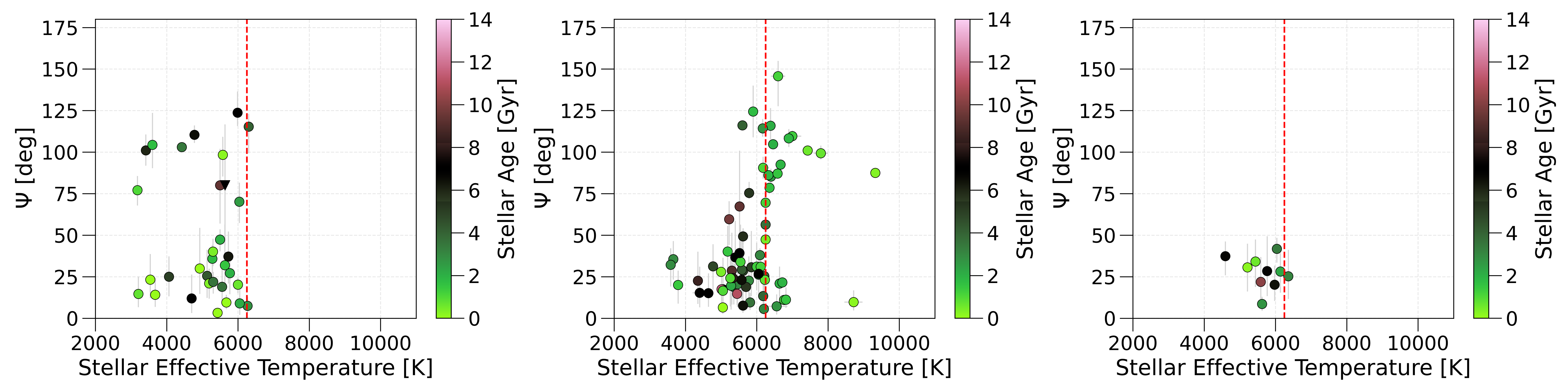}
    \caption{Relation between true obliquity $\Psi$ and stellar effective temperature $\Teff$ for different planet subgroups. The red dashed line shows the Kraft break temperature at 
    $T = 6250\,\mathrm{K}$, first highlighted by \citet{Winn_2010}. 
    In the left panel Neptunes divide almost equally between aligned and misaligned, with older planets populating even the high obliquity region. The middle and right panels show the behavior of Jupiters and super-Jupiters. The misaligned planets are younger, and they are hardly found severely before the Kraft break.}
    \label{fig:psiVSteff_planettype}
\end{figure*}

\begin{table}[h]
  \centering
  \caption{Planets with uncertain mass measurements, divided by planet type.}
  \label{tab:criticalmasses}
  \begin{tabular}{llc}
    \toprule
    Planet & $M_\mathrm{pl}$ [$\mathrm{M}_\oplus$] & Reference \\
    \midrule
    \multicolumn{3}{l}{\textbf{Neptunes:}} \\
    DS Tuc A b     & $\lesssim 14$ & 1, 2 \\
    HIP 67522 b     & $< 20$ & 3 \\
    K2-33 b        & $6.5-70$     & 4, 5 \\
    K2-290 b       & $<21.1$          & 6 \\
    Kepler-63 b    & $<120 \,(3\sigma)$  & 7 \\
    TOI-942 b      & $<14$         & 8, 9 \\
    \midrule
    \multicolumn{3}{l}{\textbf{Jupiters:}} \\
    KELT-20 b        & $<1075$     & 10 \\
    TOI-4641 b       & $<1230$          & 11 \\
    \bottomrule
  \end{tabular}

  \vspace{3pt}
  \textbf{References.} (1) \citet{Netwon_2019}, (2) \citet{Benatti_2021}, (3) \citet{Thao_2024}, (4) \citet{David_2016}, (5) \citet{Klein_2020}, (6) \citet{Hjorth_2019}, (7) \citet{Sanchis-Ojeda_2013}, (8) \citet{Zhou_2021}, (9) \citet{Carleo_2021}, (10) \citet{Lund_2017}, (11) \citet{Bieryla_2024}.
\end{table}

The small amount of rocky planets in our sample did not allow us to proceed with further analysis on them. Following \citet{Stevens_2013} classification, we focused on Neptunes ($10 - 100 \,M_{\oplus}$), Jupiters ($100 - 1000\,M_{\oplus}$) and super Jupiters ($1000-4131\,M_\oplus$). For a few planets, robust mass estimates were not available. In these cases, we consulted the literature to make an informed choice, as summarized in \cref{tab:criticalmasses} (e.g., K2-33 b, where the reported limits and simulations suggest a Neptunian nature). The result is shown in \cref{fig:psiVSteff_planettype}. 
Several authors have reported correlations between projected obliquities of hot Jupiters and stellar effective temperature \citep[e.g.,][]{Winn_2010,Albrecht_2012,Albrecht_2022}. Specifically, they observed a sudden increase in $\lambda$ beyond the Kraft break at $T_{\mathrm{eff}}\approx 6250\,\mathrm{K}$ \citep{Kraft_1967}, a threshold beyond which the convective envelope becomes nearly negligible \citep{Pinsonneault_2001}. With minor exceptions, our findings confirm the same trend in our sample for gaseous giants (\cref{fig:psiVSteff_planettype}, middle and right panel). 

Moving to Neptunes, we noted a conspicuous amount of planets on misaligned orbits, specifically 11 out of 32. This finding is in good agreement with previous works regarding the stability of misaligned --- especially polar --- orbits for Neptunes \citep{Louden_2024}. We noted that most of the old and misaligned planets in our dataset are part of the Neptunian subsample. This pattern led us to further investigate the underlying mechanisms that could produce such orbital configurations.


Some authors \citep{Batygin_2012,Lai_2014} noted that misalignment may be caused by an interaction with a binary companion. Conversely, other authors suggested that multi-star exoplanetary systems could favor aligned orbits \citep{Christian_2022,Rice_2024}. Specifically, the interaction between two moderately distant protoplanetary disks may bring to alignment.
We first examined the literature for evidence of stellar companions to the hosts of our misaligned Neptunes. Our small sample of misaligned Neptunes consisted of 11 planets. In our sample, two misaligned planets showed evidence of a binary companion: (i) WASP-131 b, whose companion has been detected with different observations \citep{Bohn_2020,Southworth_2020,Zak_2024} and (ii) K2-290, a star in a triple system orbited by two coplanar and retrograde planets \citep{Hjorth_2019,Hjorth_2021,Best_2022}. Two more aligned planets have a binary companion, namely DS Tuc A b \citep{zhou_2020} and Kepler-25 c \citep{Bourrier_2023}.

Some authors \citep{Espinoza-Retamal_2024} starting from a sample of 27 Neptune projected obliquities $\lambda$ applied a HBM framework from \citet{Dong_2023} and found tentative evidence of a dichotomous behavior in the posterior $\Psi$ distribution, although the result was defined as prior dependent. The same dichotomy was found using a subsample of 17 Neptunes with available $\Psi$ on TEPCat.

\begin{figure}[h]
    \centering
    \includegraphics[width=\linewidth]{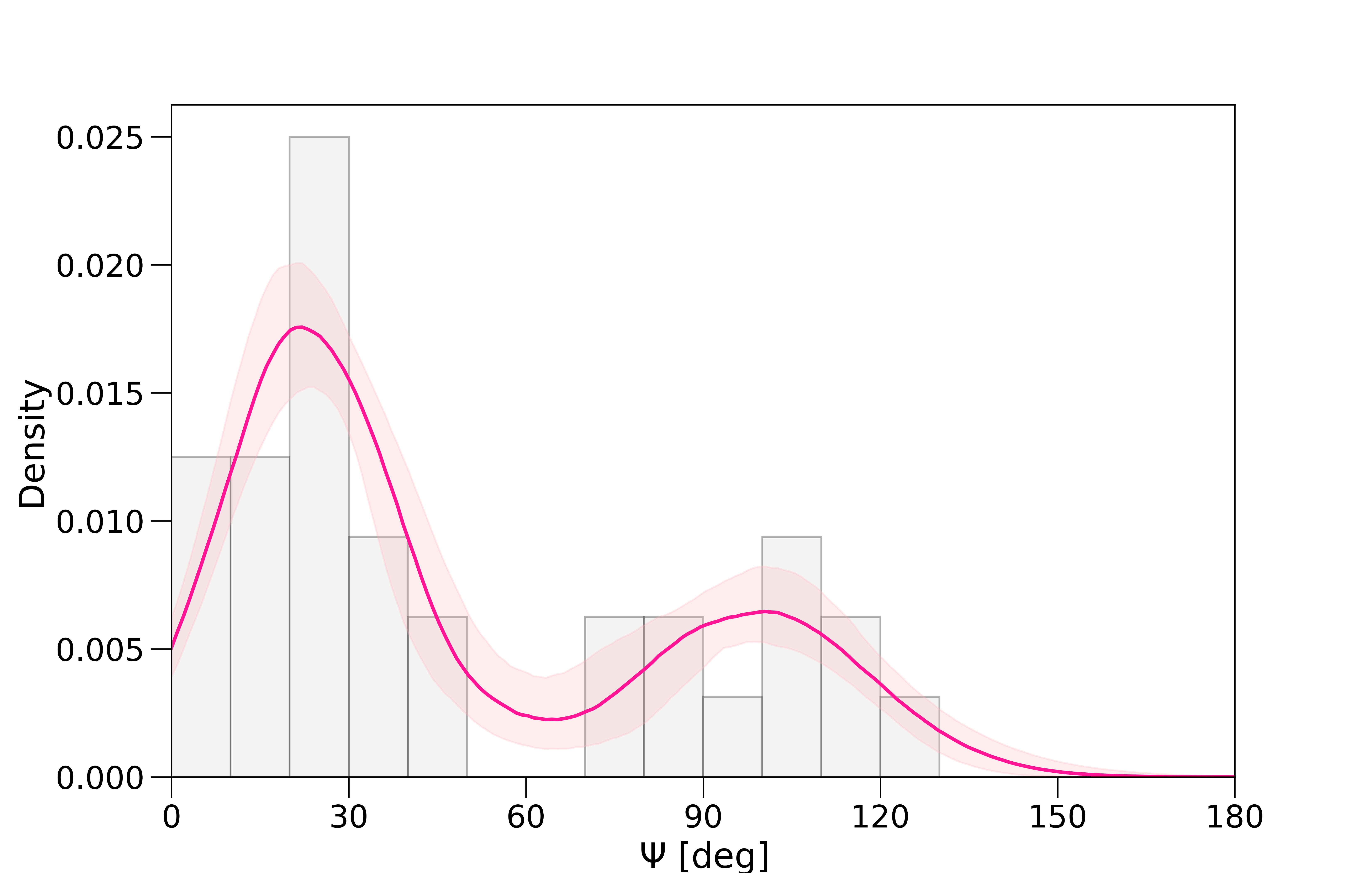}
    \caption{True obliquities distribution of the subsample of 28 Neptunes. The pink solid line represents the best fitting 2-component gaussian mixture model, obtained after 1000 perturbations of the planetary obliquities in their uncertainty range. The shaded light pink area represents the $68\%$ CI. The two components are well separated by an Ashman factor $D= 4.33.$}
    \label{fig:neptunianfit}
\end{figure}

\begin{figure}[h]
    \centering
    \includegraphics[width=\linewidth]{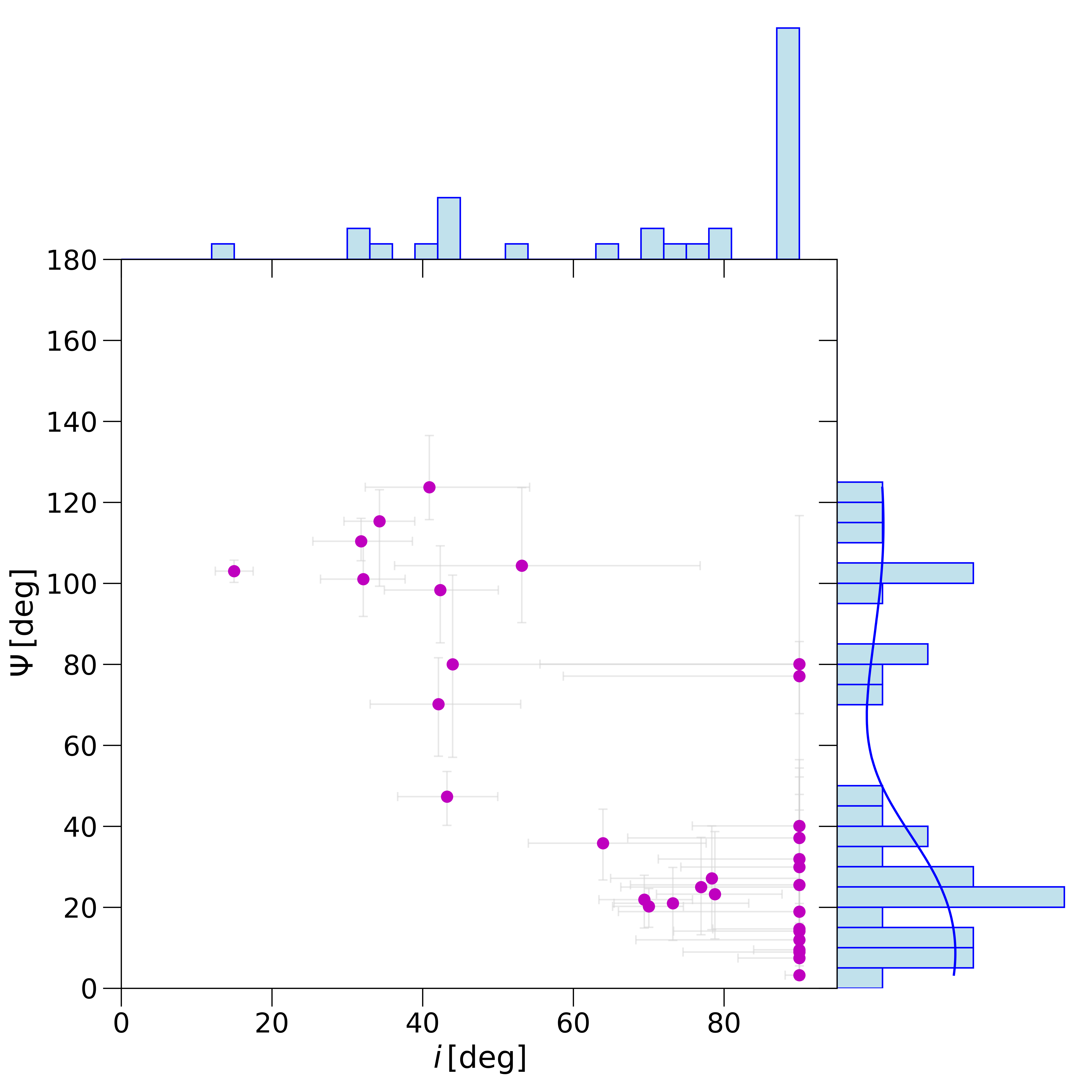}
    \caption{True spin-orbit obliquities of our sample of 32 Neptunes in respect to the stellar inclination. In this figure, following the same fashion of previous plots, we can note (i) how the dichotomous behavior of true obliquities is accompanied by a non-uniform distribution of stellar obliquities and (ii) how statistics is still low to infer robust conclusions.}
    \label{fig:psiVSistar_neptunes}
\end{figure}
We analyzed our sample of Neptunian planets perturbing their $\Psi$ values in their uncertainty bounds, using split-normal distributions, for 1000 times. We fitted each synthetic sample with a two-component gaussian distribution and we extracted the median values for the parameters of the gaussian distributions $\phi_i=\mathcal{N}(\mu_i, \sigma_i)$ and their relative weights $w$, as well as the 16th and 84th percentile. The result, shown in \cref{fig:neptunianfit}, suggests the presence of two main components $\phi_1=\mathcal{N}(23.21^{+3.37}_{-3.73},14.49^{+2.82}_{-2.74})$ and $\phi_2=\mathcal{N}(98.66^{+6.79}_{-8.21}, 20.02^{+6.68}_{-5.41})$ of relative weights $w_1 = 0.66\pm0.06$ and $w_2 = 0.34\pm 0.06$, furthermore supported by an Ashman value of $D=4.33^{+0.82}_{-0.97}$, well beyond the $D\gtrsim 2$ rule-of-thumb indicating that the two components are clearly separated \citep{Ashman_1994}. However a diptest conducted on this sample was unable to reject the unimodality of the distribution. Thus, these tests mildly corroborate the existence of a dichotomy in the largest Neptunian $\Psi$ sample available as of today, although the sample size is still to low to infer anything but tentative conclusions. Moreover the stellar inclination distribution of Neptunian planets show clear signs of anisotropy (\cref{fig:psiVSistar_neptunes}), while both the results and marginal distributions of the whole sample without the Neptunes remain unchanged.

A larger sample of Neptunes is surely needed to extract definitive conclusions regarding their distributions in the $\Psi$ space.
\section{Discussion and conclusions\label{sec:discussion}}

Starting from a sample of 264 projected spin-orbit obliquities $\lambda$ collected from TEPCat, the NASA Exoplanet Archive, and additional literature sources, we were able to homogeneously compute 116 true obliquities $\Psi$ through the rotation period method. Together with four further values taken directly from the literature, and after excluding those derived via gravity darkening (GD) due to its known bias towards perpendicular orbits \citep{Siegel_2023}, this results in a final de-biased sample of 120 planets. This dataset doubles the size of the previous compilation by \citet{Albrecht_2021} and constitutes the largest collection of true spin-orbit obliquities available to date. The main results emerging from the analysis of this enlarged sample are outlined in the following subsections.

\subsection{No clustering of misaligned orbits}
Our primary goal was to shed light on the true spin-orbit obliquity distribution of exoplanets using a data-driven approach. This was achieved by correctly inferring stellar inclinations $\istar$ \citep{Masuda_2020,Bowler_2023,Morgan_2024}, which allowed us to extract the true obliquities following \cref{eq:cospsi} \citep[e.g.,][]{Fabrycky_2009,Albrecht_2021,Knudstrup_2024}. \citet{Albrecht_2021} noted a dichotomy in the distribution of $\Psi$, with two distinct clusters roughly within the ranges $\Psi=0^\circ-41^\circ$ and $\Psi = 80^\circ-125^\circ$. This conclusion, which relied on still uncertain theoretical foundation \citep{Albrecht_2021}, has been questioned by some teams \citep{Siegel_2023,Dong_2023}, which subsequent work has attempted —-- so far unsuccessfully --— to resolve it \citep{Knudstrup_2024}. However, \citet{Siegel_2023} and \citet{Knudstrup_2024} pointed that stellar inclination $\istar$ could significantly bias the studied sample.

In this work, we provide strong evidence against the existence of two peaks for aligned and perpendicular orbits, finally supporting with observational data the idea of a single peak for equatorial planets followed by an isotropic distribution on misaligned obliquities \citep{Siegel_2023,Dong_2023}. This is granted by a dip-test rejecting the null hypothesis of uni-modality at the $1\%$ level. Although this conclusion was inferred without relying on GD measurements, future works --- based on a significantly larger number of GD determinations when available --- could provide interesting insights into the impact of this technique on the overall exoplanets sample from a data-driven perspective.

\subsection{Stellar inclination bias}
In particular, in this work we show how $\istar$ can heavily bias the $\Psi$ distribution: RM measurements are performed during transits, so configurations with $\istar\sim 0^\circ$ are more likely to host misaligned planets, whereas configurations with $\istar\sim 90^\circ$ favor aligned orbits. Unlike \citet{Albrecht_2021}, our sample is characterized by more uniform stellar inclinations. Even a small non-uniformity, such as removing inclinations lower than $35^\circ$, can recreate the apparent dichotomy observed in the aforementioned work. This corroborates from a data-driven perspective a bias that was theorized in previous research \citep{Siegel_2023,Dong_2023}.

\subsection{A tentative Neptunian dichotomy}

Our 120-planets sample presents a sub-sample of 32 planets belonging to the Neptunian realm, according to \citet{Stevens_2013} classification ($10 < M_{\mathrm{nep}}/M_\oplus < 100$). Among these, 11 show signs of misalignment, defined as being more than $1\sigma$ from $41^\circ$, limit beyond which ZKL cycles can shape the planetary system architecture in case of a massive companion. In this regard, our analysis found evidence of a stellar companion only for four Neptunians, detailed in \cref{subsec:psiage}, among which two show signs of misalignment. Previous authors \citep{Petrovich_2020, Espinoza-Retamal_2024} have found tentative evidence of a dichotomous behavior of the true obliquity distribution of Neptunes, basing their work on projected obliquities $\lambda$. With a larger sample of true obliquities we can robustly state that a multi-Gaussian fit yields two distributions peaked at $\sim 20^\circ$ and $\sim 100^\circ$ and clearly separated by an Ashman $D$ value of $4.33$ after a 2-Gaussian fit. However, a Hartigan's dip-test \citep{diptest} is not able to reject the hypothesis of uni-modality for this distribution, probably because of the still small sample.

We cannot help but notice that this dichotomy very much resembles what \citet{Albrecht_2021} found for the whole sample of exoplanets. A Neptunian sample half the size of the one used by \citet{Albrecht_2021}, combined with a failed dip-test might suggest that a larger sample is needed to infer definitive conclusions, as low statistics can hinder the reliability of p-values by masking effects that a larger sample will show. This conclusion agrees with the results of \citet{Espinoza-Retamal_2024}, which detailed how the posterior distribution of Neptunian true obliquities is sensitive to hyperpriors in \citet{Dong_2023} hierarchical Bayesian framework, whereas Jupiters tend to cluster on aligned orbits with an isotropic tail on larger obliquities without strongly responding, and modifying the overall trend, to hyperpriors. 

\citet{Albrecht_2021} noted how theoretical foundations lacked to justify their reported dichotomy, however they highlighted how secular resonance crossing in presence of a massive outer companion could provide theoretical support for the presence specifically of perpendicular Neptunes, similarly to ZKL cycles \citep{Petrovich_2020}. In our sample only one planet presents an outer companion while being misaligned, hence we can just suggest this as possible mechanism for misalignment. Several mechanisms beyond ZKL have been proposed to explain misalignment, such as primordial misalignment \citep{Bate_2010} and magnetic torques \citep{Lai_2011} (see \citet{Albrecht_2022} for a review). Other works have highlighted that when specific conditions occur Neptunians can show peculiar architectures: compact Neptunian systems tend to be more aligned \citep{Razdom_2024,Polanski_2025}, Neptunians are less responsive to tides, hence their longer realignment timescales can doom the planet to reside in a misaligned orbit \citep{Handley_2024}, and polar Neptunes can be so unresponsive to tides to retain their perpendicular obliquity \citep{Louden_2024}. 

The measurement of spin–orbit angles for Neptune-sized planets is particularly valuable, as their obliquities may provide crucial clues on the dynamical histories of intermediate-mass planets, a still underrepresented population that appears to show intriguing differences from hot Jupiters. The ongoing interest for Neptunes, along with future mission to characterize small sized planets \citep[PLATO][]{Rauer_2025} and exoplanetary atmospheres \citep[ARIEL][]{Tinetti_2018}, will dramatically increase our knowledge of formation and evolution of exoplanets with the fresh data they are expected to produce in the next years.


\begin{acknowledgements}
We deeply thank the referee for their constructive and insightful comments, which greatly contributed in improving the quality of this work.
AMR gratefully thanks M. Frangiamore for the enlightening conversations during these months.
FB acknowledges support from Bando Ricerca Fondamentale INAF 2023 and from PLATO ASI-INAF agreement n. 2022-28-HH.0.
SF acknowledges financial contribution from the European Union (ERC, UNVEIL, 101076613), and from PRIN-MUR 2022YP5ACE. Views and opinions expressed, however, are those of the author(s) only and do not necessarily reflect those of the European Union or the ERC. Neither the European Union nor the granting authority can be held responsible for them.
\\
This work has made an extensive use of \hyperlink{https://www.astro.keele.ac.uk/jkt/tepcat/}{TEPCat} by John Southworth, Keele University, UK. This research has made use of the SIMBAD database, operated at CDS, Strasbourg, France \citep{Wenger_2000}. This research has made use of the \href{https://exoplanetarchive.ipac.caltech.edu/}{NASA Exoplanet Archive}, which is operated by the California Institute of Technology, under contract with the National Aeronautics and Space Administration under the Exoplanet Exploration Program \citep{NEA}. This research has made use of data obtained from or tools provided by the portal \href{https://exoplanet.eu}{exoplanet.eu} of The Extrasolar Planets Encyclopedia. This work has made use of data from the European Space Agency (ESA) mission Gaia (\url{https://www.cosmos.esa.int/gaia}), processed by the Gaia Data Processing and Analysis Consortium (DPAC, \url{https://www.cosmos.esa.int/web/ gaia/dpac/consortium}). Funding for the DPAC has been provided by national institutions, in particular the institutions participating in the Gaia Multilateral Agreement \citep{Gaia}. This research has made use of the Astrophysics Data System, funded by NASA under Cooperative Agreement 80NSSC21M00561.
\\
\textit{Software.} The code associated with this work used only open source software. This research made use of \textsc{astroquery} \citep{astroquery}, \textsc{emcee} \citep{emcee}, \textsc{corner.py} \citep{corner} \textsc{Matplotlib} \citep{matplotlib}, \textsc{NumPy} \citep{numpy}, \textsc{pandas} \citep{pandas}, \textsc{SciPy} \citep{scipy}, \textsc{schwimmbad} \citep{schwimmbad}, \textsc{seaborn} \citep{seaborn}.

\end{acknowledgements}


\bibliographystyle{aa} 
\bibliography{bibliography.bib}

\begin{sidewaystable*}

\setlength{\tabcolsep}{0.13cm} 
\renewcommand{\arraystretch}{1.5} 
\footnotesize
\begin{tabularx}{\linewidth}{l c c c c c c c c c c c}
  \toprule
  Planet & $\Teff$ & $M_p$ & $\rstar$ & $\vsini$ & $\prot$ & \iorb & $\lambda$ & $\Psi_\mathrm{literature}$ & $\istar$ & $\Psi$ & Age \\ 
         & [K] & $[\Mearth]$ & $[\Rsun]$ & $[\mathrm{km\,s^{-1}}]$ & [days] & [deg] & [deg] & [deg] & [deg] & [deg] & [Gyr] \\ 
  (1) & (2) & (3) & (4) & (5) & (6) & (7) & (8) & (9) & (10) & (11) & (12) \\ 
  \midrule
    55 Cnc e & $5172 \pm 18$ [1] & $7.99^{+0.32}_{-0.33}$ [2] & $0.974^{+0.004}_{-0.001}$  & $2.00^{+0.43}_{-0.47}$ [1] & $38.8 \pm 0.05$ [1] & $83.9^{+0.6}_{-0.5}$ [1] & $10.0^{+17.0}_{-20.0}$ [1] & $23.0^{+14.0}_{-12.0}$ [1] & $90.0^{+0.0}_{-23.2}$  & $21.99^{+14.45}_{-11.98}$  & $10.2 \pm 2.5$ [3] \\
    AU Mic b & $3540^{+120}_{-110}$ [4] & $10.2^{+3.9}_{-2.7}$ [5] & $0.74 \pm 0.02$ [6] & $9.23^{+0.79}_{-0.31}$ [7] & $4.9 \pm 0.7$ [8] & $89.58 \pm 0.38$ [6] & $-4.7^{+6.8}_{-6.4}$ [7] & $...$  & $90.0^{+0.0}_{-16.7}$  & $14.11^{+10.13}_{-7.3}$  & $0.020^{+0.003}_{-0.002}$ [6] \\
    CoRoT-2 b & $5625 \pm 120$ [9] & $1102 \pm 70$ [9] & $0.881 \pm 0.003$  & $11.85 \pm 0.50$ [9] & $4.52 \pm 0.02$ [10] & $87.8 \pm 0.1$ [11] & $7.2 \pm 4.5$ [9] & $...$  & $90.0^{+0.0}_{-6.3}$  & $8.6^{+4.4}_{-4.1}$  & $2.7^{+3.2}_{-2.7}$ [12] \\
    ... & ... & ... & ... & ... & ... & ... & ... & ... & ... & ... & ... \\
    ... & ... & ... & ... & ... & ... & ... & ... & ... & ... & ... & ... \\
  \bottomrule
\end{tabularx}
\caption{Extract from our dataset. The full dataset is available at CDS as a machine-readable table.\\ 
\textbf{Note.} Column (1): planet name. Column (2): stellar effective temperature. Column (3): planet mass. Column (4): stellar radius. Column (5): stellar projected rotational velocity. Column (6): stellar rotation period. Column (7): planet's orbital inclination. Column (8): projected spin-orbit obliquity. Column (9): true spin-orbit obliquity, if available in the literature. Column (10): stellar spin inclination. Column (11): true spin-orbit angle computed in this work. Column (12): stellar age. Stellar radii $\rstar$ are taken from Gaia DR3 \citep{GaiaDR3}, unless otherwise specified.
\\ \textbf{References.} Stellar inclinations $\istar$ and true obliquities $\Psi$ are derived in this work. The remaining quantities are taken from: 
[1] \citet{Zhao_2023},
[2] \citet{Bourrier_2018a},
[3] \citet{vonBraun_2011},
[4] TEPCat \citet{Southworth_2011},
[5] \citet{Donati_2023},
[6] \citet{Wittrock_2023},
[7] \citet{Hirano_2020a},
[8] \citet{Zuniga-Fernandez_2020},
[9] \citet{Bouchy_2008},
[10] \citet{Bonomo_2017},
[11] \citet{Alonso_2008},
[12] \citet{Gillon_2010}.}

\label{tab:data}
\end{sidewaystable*}

\end{document}